\pdfoutput=1

\documentclass[11pt]{elsarticle}

\makeatletter
\def\ps@pprintTitle{%
  \let\@oddhead\@empty
  \let\@evenhead\@empty
  \let\@oddfoot\@empty
  \let\@evenfoot\@oddfoot
}
\makeatother

\usepackage{url}
\usepackage{breakurl}
\usepackage[breaklinks,
            colorlinks = true,
            linkcolor = blue,
            urlcolor  = blue,
            citecolor = blue,
            anchorcolor = blue]{hyperref}

\usepackage{lineno}
\usepackage{graphicx}
\usepackage[margin=1.25in]{geometry}
\usepackage[usenames,dvipsnames]{color}

\newcommand\acp{\begin{center}
\rule[-0.2in]{\hsize}{0.01in}\\\rule{\hsize}{0.01in}\\
\vskip 0.1in Submitted to the  Proceedings\\ 
of the African Conference on Fundamental and Applied Physics
    \vskip 0.05in
    {\it Second Edition, ACP2021, March 7--11, 2022 --- Virtual Event}\\
\rule{\hsize}{0.01in}\\\rule[+0.2in]{\hsize}{0.01in} \\
\end{center}}

\usepackage[firstpage=true]{background}
\backgroundsetup{contents={\parbox{6.5in}{\acp}}, scale=1,placement=top,opacity=1,color=black,position={3.25in,1.2in}}

\usepackage{fancyhdr}
\fancypagestyle{plain}{%
  \fancyhf{}%
  \fancyhead[C]{}
  \fancyfoot[C]{\thepage}
}

\fancypagestyle{empty}{%
  \fancyhf{}%
  \fancyhead[C]{{\it ACP2021, March 7--11, 2022 --- Virtual Edition}}
  \fancyfoot[C]{\thepage}
}
\begin{document}

\begin{frontmatter}


\title{Search for invisible Higgs bosons produced via vector boson fusion at the LHC using the ATLAS detector}

\author[add1]{Mohamed Zaazoua\corref{cor1}}
\ead{mohamed.zaazoua@cern.ch}
\author[add1]{Farida Fassi}
\author[add2]{Kétévi Adiklè Assamagan}
\author[add2]{Diallo Boye}

\cortext[cor1]{Corresponding Author}

\address[add1]{Mohammed V University in Rabat, Morocco}
\address[add2]{Brookhaven National Laboratory, USA}

\begin{abstract}
\noindent 
Despite dark matter abundance, its nature remains elusive.
Many searches of dark matter particles are carried out using different technologies either via direct detection, indirect detection, or collider searches. 
In this work, the invisible Higgs sector was investigated, where Higgs bosons are produced via the vector boson fusion (VBF) process and subsequently decay into invisible particles.
The hypothesis under consideration is that the Higgs boson might decay into a pair of weakly interacting massive particles (WIMPs), which are candidates for dark matter.
The observed number of events are found to be in agreement with the background expectation from Standard Model (SM). Assuming a 125 GeV Higgs boson with SM production cross section, the observed and expected upper limits on the branching fraction of its decay into invisible particles are found to be 0.13 at 95\% confidence level.
Combination of searches for an invisibly decaying Higgs boson produced via the main Higgs production modes at the LHC using 2011–2018 data is conducted and discussed. 
\end{abstract}


\end{frontmatter}

%



\section{Introduction}
\label{sec:intro}
\noindent
Based on many astrophysical observations, there are strong evidences that dark matter exists and makes up most of matter in the universe. 
Yet dark matter is completely invisible to traditional detectors, and its nature remains an open question; we can only infer existence through its gravitational effects. 
Higgs portal model is considered as one of the important paths for dark matter searches, where the Higgs could be a mediator between SM particles and ones that belong to the dark sector.

The analysis uses data samples produced with a luminosity of 139 $\textrm{fb}^{-1}$ of proton-proton collisions at center of mass of $\sqrt{s}=13~$TeV, recorded by the ATLAS detector at the LHC. 
The SM predicts that no more than 0.1\% of the branching fraction ($B_{H \rightarrow inv}$) of the Higgs bosons goes into invisible parrticles, $H\rightarrow ZZ^{*}\rightarrow4\nu$. 
Beyond Standard Model scenarios predict larger values for the invisible Higgs branching ratio, up to 10\% \cite{B_langer_2013}. 
%
The choice of VBF topology offers a powerful rejection against background.
The contribution of the gluon fusion (ggF) process in the signal region (SR) is relatively small compared to that of the VBF process, but it is also considered to be part of the signal.
The experimental signature of VBF is two leading jets with a large rapidity gap ($\Delta\eta_{jj}$) and large invariant mass $m_{jj}$.
The events entering the SR are required to have large missing energy transverse.
Events with two, three or four jets are considered as well. 
To benefit from the large data set, a binning on $m_{jj}$ is applied (5 bins), and for the first time, a binning on $\Delta\Phi_{jj}$ (2 bins) and an additional binning for the jet multiplicity, resulting in a total of 11 bins. 
%
%
The signal and control regions (CR) are defined using selections that exploit the four leading jets, the leptons, and the missing transverse energy ($E^{miss}_{T}$).
The analysis defines two control regions to estimate the V+jets background; Z($\rightarrow ll$) and W($\rightarrow l \nu $) to estimate the contribution of  Z($\rightarrow \nu \nu $) and W($\rightarrow l_{lost} \nu $) in the SR. 
\section{Background estimation}
\begin{figure}[!htbp]
\begin{center}
\includegraphics[width=0.39\textwidth]{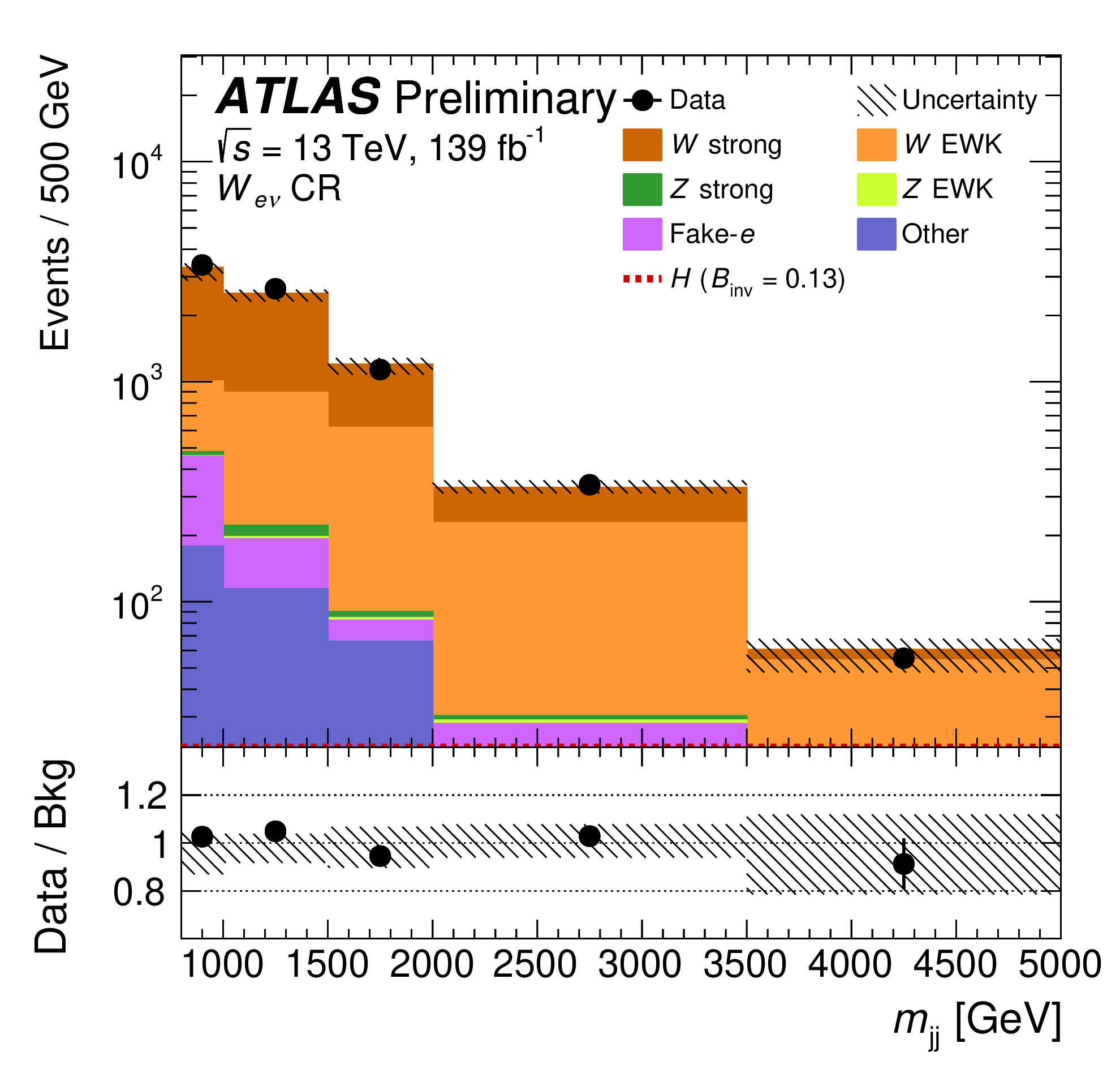}
\includegraphics[width=0.39\textwidth]{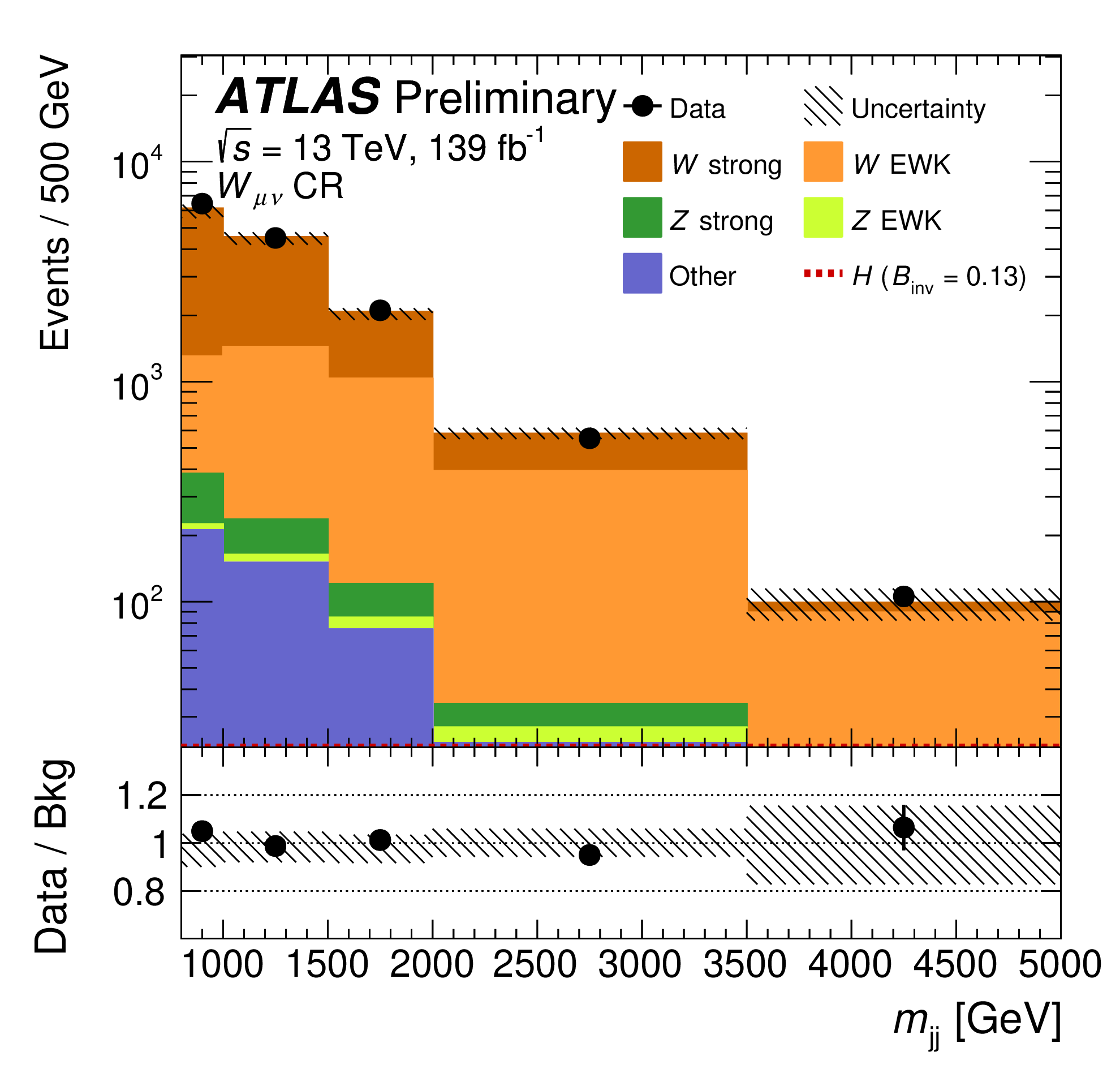}
\end{center}
\caption{
Di-jet invariant mass distribution in 
$W\rightarrow e\nu$ and $W\rightarrow\mu\nu$ control regions. Statistical and reconstruction systematic uncertainties are presented with the hashed band \cite{vbfrun--2}.}
\label{fig:Zbckg}
\end{figure}

\begin{figure}[!htbp]
\begin{center}
\includegraphics[width=0.39\textwidth]{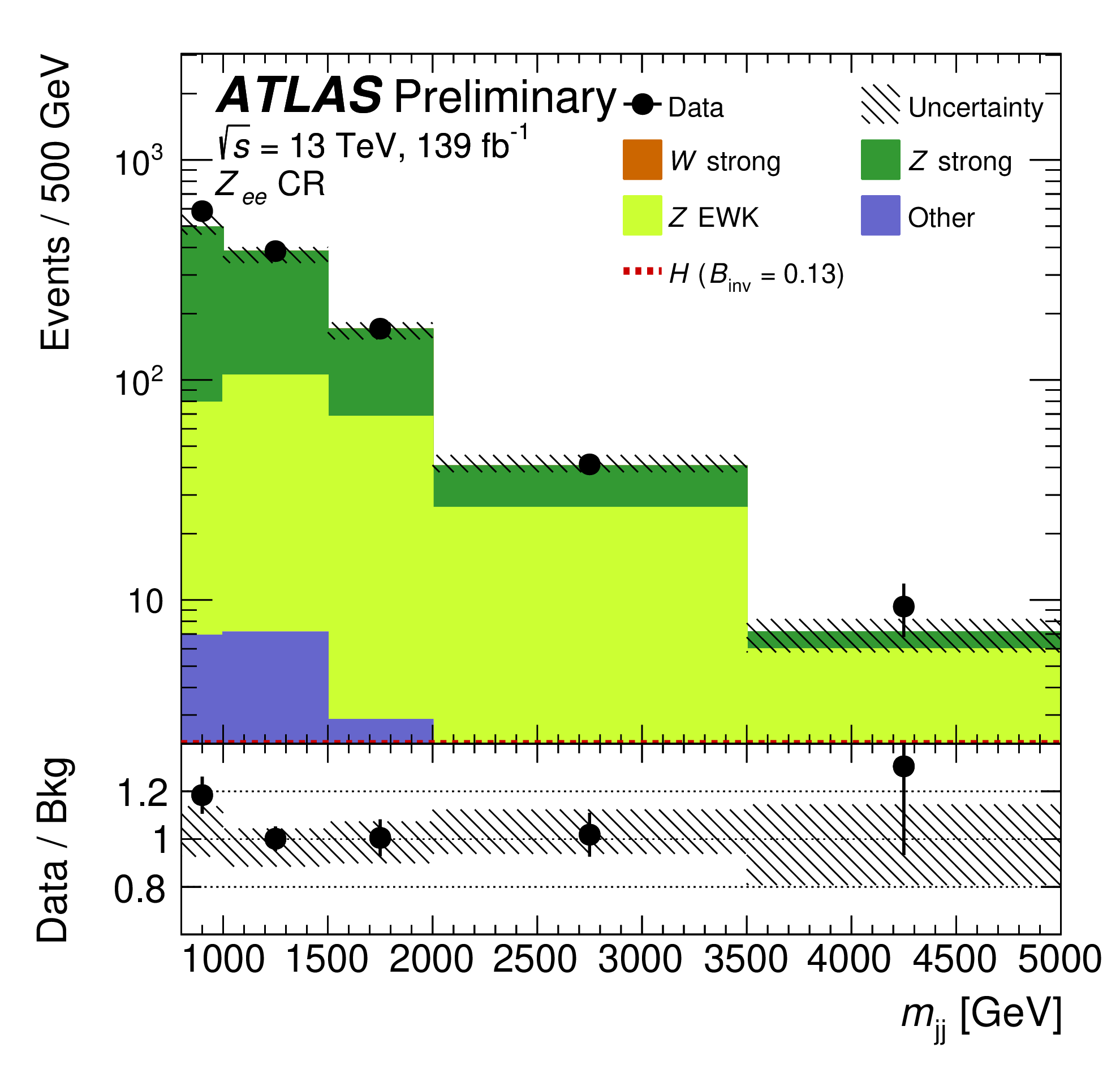}
\includegraphics[width=0.39\textwidth]{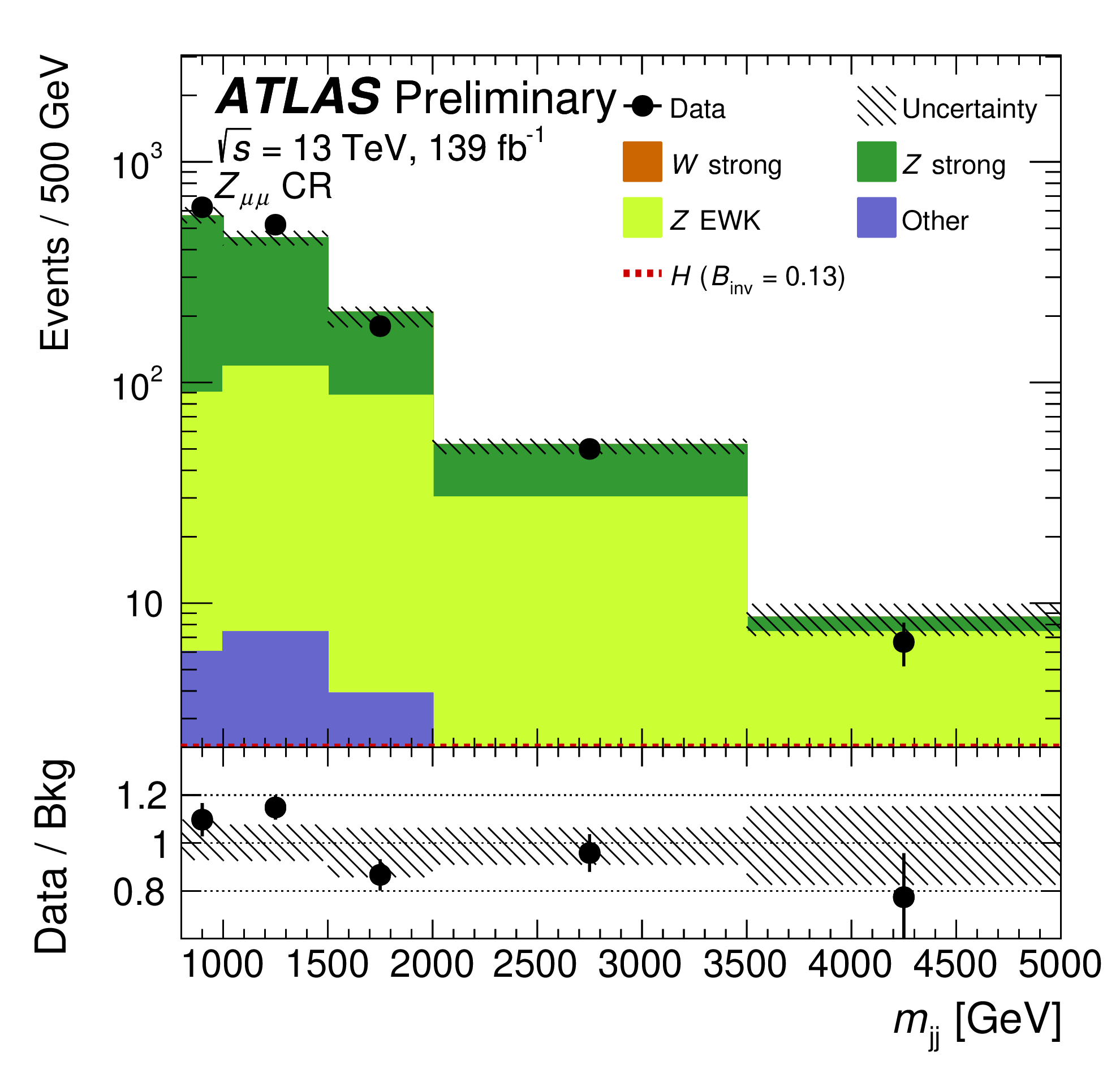}
\end{center}
\caption{
Di-jet invariant mass distribution in  $Z\rightarrow ee$ and $Z\rightarrow\mu\mu$ control regions. Statistical and reconstruction systematic uncertainties are presented with the hashed band \cite{vbfrun--2}.}
\label{fig:Wbckg}
\end{figure}

The V+jets events represent the dominant background $\sim$ 95\% in the SR. There are other less important sources of background, such as diboson, $t\bar{t}$, and multijet (MJ) processes. 
To constrain the V+jets backgrounds, the analysis uses a data-driven technique based on W and Z CR that are divided into 11 bins.  
In both the cases of W and Z CR, the simulation is normalized to cross section times luminosity as shown in Figures \ref{fig:Zbckg}  and \ref{fig:Wbckg}.
 W CR ($W^{high}_{e\nu}$); which are enriched by events passing the missing transverse momentum significance requirement; can have a small contribution from events with electrons faking jets. 
To estimate this contribution, the analysis defines a further $W^{low}_{e\nu}$ CR in which events that fail the $E^{miss}_{T}$ significance cut are selected, with a loose identification requirement for the electrons. 

Events from the multijet background are characterized by a large $\Delta\Phi_{jj}$ between the two leading jets and a small value of $E^{miss}_{T}$ from mis-measurement of jets.
Its contribution is estimated from the data using the rebalance and smear (R\&S) technique \cite{vbfrun--2}, which uses an inclusive jet sample recorded by a single jet trigger as input.
Performance of the method is validated in low $E^{miss}_{T}$  and low $m_{jj}$ CR.

\section{{ Systematic uncertainties}}
The high-order matrix element effects and parton shower matching uncertainties are evaluated using the  renormalisation, factorisation, resumation (also called "qsf") and CKKW matching scales.
The renormalization and factorization scales are varied up and down by a factor of two, using on-the-fly event weights in the Sherpa MC samples \cite{vbfrun--2}. 
The uncertainties are calculated as relative error for the CKKW matching scales.
The PDF uncertainties defined as the standard deviation of the ensemble of 100 PDFs within the NNPDF set \cite{vbfrun--2}. 
Three main categories of experimental uncertainties affect the sensitivity of the analysis: uncertainty on the luminosity, the trigger efficiency, and uncertainties related to the used physics objects.

\section{{ Results}}
\begin{figure}[!htbp]
\begin{center}
\includegraphics[width=0.75\textwidth]{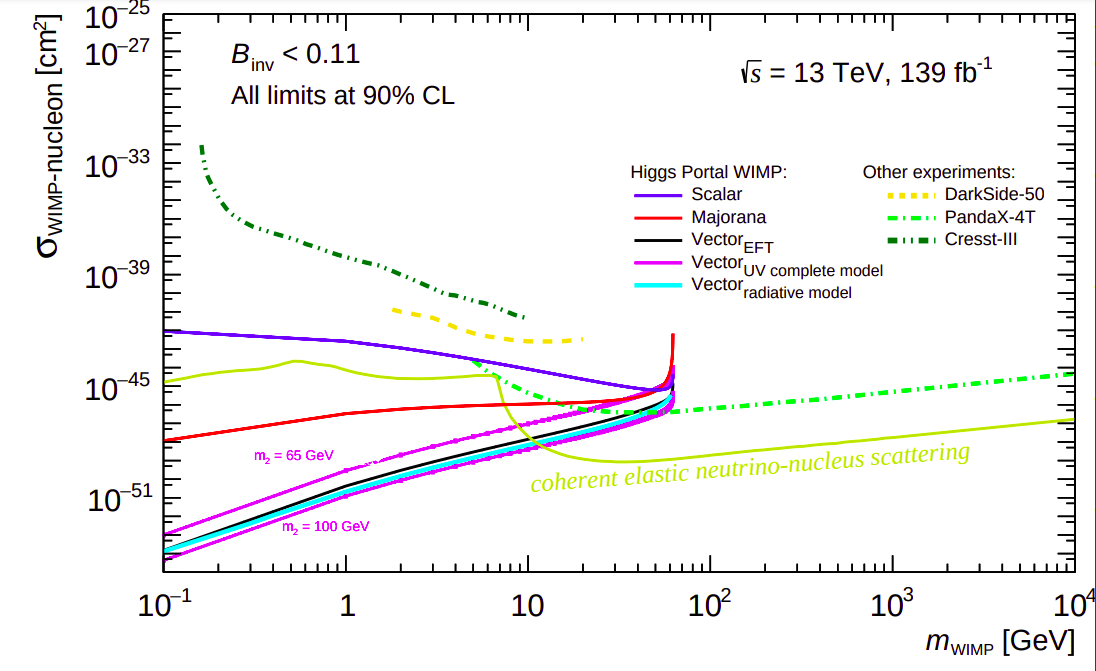}
\end{center}
\caption{ Upper limits on the spin independent WIMP-nucleon cross section using Higgs portal interpretation of invisible branching fraction at $90\%$  CL as. a function of the WIMP mass. For the vector-like WIMP hypothesis, the dependence on mass $m_2$  of the new scalar particle, which is often predicted by renormalizable models, is shown for three different values covering a wide range \cite{VectorDM}. For comparison with direct searches for DM, the graph shows results from Refs.\cite{VectorDM,vbfrun--2paper}. 
The neutrino floor for coherent elastic neutrino-nucleus scattering assumes germanium as the target over the whole WIMP mass range
    \cite{VectorDM,vbfrun--2paper}.}
\label{fig:overlay}
\end{figure}

The event yields in the eleven bins  of all the signal and control regions  after the likelihood fit are found to be in a good agreement with the expected yields from the SM background. 
Since no excess of events was found, a limit setting on the invisible Higgs branching fraction gives a more stringent value of 13\% at 95\% confidence level (CL). 
The latest results of this analysis show improvement and were used for a Higgs-portal dark matter interpretation to sub-GeV regime and with the addition of vector WIMP.  \cite{vbfrun--2paper,VectorDM} as shown in Figure \ref{fig:overlay}.

\section{Conclusions}
\label{sec:conc}
This work presents a search for Higgs boson produced via VBF mechanism and decay into invisible particles using a luminosity of 139 $\textrm{fb}^{-1}$ of proton-proton collision at center of mass $\sqrt{s}=$13 TeV collected during the Run2 period using  the ATLAS detector at the LHC. 
The experimental signature of the VBF process is two leading jets with a large rapidity gap ($\Delta\eta_{jj}$) and large invariant mass $m_{jj}$ and $E^{miss}_{T}$. 
Events with three or four jets are considered as well if they originate from initial or final state radiations. An upper limit of 0.13 at 95\% CL is set on $B_{H \rightarrow inv}$ assuming the SM Higgs boson production. 
This result is used in the Higg-portal dark matter interpretation to set a limit on the WIMP-nucleon cross section as a function of the WIMP mass.



\bibliographystyle{elsarticle-num}
\bibliography{myreferences.bib}

\begin{thebibliography}{1}
\expandafter\ifx\csname url\endcsname\relax
  \def\url#1{\texttt{#1}}\fi
\expandafter\ifx\csname urlprefix\endcsname\relax\def\urlprefix{URL }\fi
\expandafter\ifx\csname href\endcsname\relax
  \def\href#1#2{#2} \def\path#1{#1}\fi

\bibitem{B_langer_2013}
G.~B{\'{e}}langer, B.~Dumont, U.~Ellwanger, J.~Gunion, S.~Kraml,
  \href{https://doi.org/10.1016%2Fj.physletb.2013.05.024}{Status of invisible
  higgs decays}, Physics Letters B 723~(4-5) (2013) 340--347.
\newblock \href {https://doi.org/10.1016/j.physletb.2013.05.024}
  {\path{doi:10.1016/j.physletb.2013.05.024}}.
\newline\urlprefix\url{https://doi.org/10.1016%2Fj.physletb.2013.05.024}

\bibitem{vbfrun--2}
\href{https://inspirehep.net/literature/1791623}{{Search for invisible Higgs
  boson decays with vector boson fusion signatures with the ATLAS detector
  using an integrated luminosity of 139 fb$^{-1}$}} (4 2020).
\newline\urlprefix\url{https://inspirehep.net/literature/1791623}

\bibitem{vbfrun--2paper}
{ATLAS Collaboration}, \href{https://arxiv.org/abs/2202.07953}{Search for
  invisible higgs-boson decays in events with vector-boson fusion signatures
  using 139 $\textrm{fb}^{-1}$ of proton-proton data recorded by the atlas
  experiment} (2022).
\newblock \href {https://doi.org/10.48550/ARXIV.2202.07953}
  {\path{doi:10.48550/ARXIV.2202.07953}}.
\newline\urlprefix\url{https://arxiv.org/abs/2202.07953}

\bibitem{VectorDM}
M.~Zaazoua, L.~Truong, K.~A. Assamagan, F.~Fassi,
  \href{https://arxiv.org/abs/2107.01252}{Higgs portal vector dark matter
  interpretation: review of effective field theory approach and ultraviolet
  complete models} (2021).
\newblock \href {https://doi.org/https://doi.org/10.31526/lhep.2022.270}
  {\path{doi:https://doi.org/10.31526/lhep.2022.270}}.
\newline\urlprefix\url{https://arxiv.org/abs/2107.01252}

\end{thebibliography}

\end{document}